\documentclass{PoS}

\title{Non-perturbative renormalization of the energy-momentum tensor in SU(3) Yang-Mills theory}

\ShortTitle{Non-perturbative renormalization of the energy-momentum tensor in SU(3)
Yang-Mills theory}

\author{\speaker{M. Pepe}\\
        INFN, Sezione di Milano-Bicocca\\ 
        Edificio U2, Piazza della Scienza 3\\ 
        20126 Milano, Italy.\\
        E-mail: \email{Michele.Pepe@mib.infn.it}}
\author{Leonardo Giusti\\
        Dipartimento di Fisica, Universit\`a di Milano-Bicocca\\
        and INFN, sezione di Milano-Bicocca\\
        Edificio U2, Piazza della Scienza 3\\ 
        20126 Milano, Italy.\\
       E-mail: \email{Leonardo.Giusti@mib.infn.it}}

     \abstract{We present a strategy for a non-perturbative determination of the finite
       renormalization constants of the energy-momentum tensor in the SU(3) Yang-Mills
       theory. The computation is performed by imposing on the lattice suitable Ward
       Identites at finite temperature in presence of shifted boundary conditions. We show
       accurate preliminary numerical data for values of the bare coupling $g_0^2$ ranging
       for 0 to 1.}

\FullConference{The 32nd International Symposium on Lattice Field Theory\\
                 23-28 June, 2014\\
                 Columbia University New York, NY}

\begin{document}

\section{Introduction}
The lattice regularization breaks the space-time symmetries of translations and
rotations down to discrete subgroups. The full group is then recovered in the limit of vanishing
lattice spacing. The energy-momentum tensor $T_{\mu\nu}$ is a field that contains crucial 
information about the theory. The spatial integral of $T_{00}$ is the energy of the system,
the integral of the $T_{0k}$ component is the charge of translations in the spatial direction 
$\hat k$, and the combination $\varepsilon_{ijk}x_i T_{0j}$ corresponds to rotations 
around $\hat k$. Furthermore, $T_{\mu\nu}$ provides information about the thermodynamics of 
the quantum theory at finite temperature. The $\langle T_{00} \rangle$ and $\langle T_{kk}\rangle$ expectation values
represent the energy density and the pressure, respectively. In a moving reference frame,
the entropy can be obtained from $\langle T_{0k}\rangle$~\cite{Landau}.

At fixed lattice spacing the energy-momentum tensor can be defined in many different
ways. After renormalization all definitions differ by irrelevant terms that give vanishing
contributions in the continuum limit, and the correct Ward identities of translations
and rotations are properly recovered~\cite{Caracciolo89}. The traceless components of $T_{\mu\nu}$
pick up ultraviolet finite multiplicative renormalization factors that approaches 1
as the bare coupling constant $g_0 \rightarrow 0$.

The non-perturbative computation of the renormalization factors is a necessary step to perform the continuum limit
extrapolation of correlators of the energy-momentum tensor obtained at finite lattice spacing. In this talk 
we consider the $SU(3)$ Yang-Mills theory on the lattice in four dimensions, and we present an efficient technique 
to determine non-perturbatively the renormalization constant $Z_T(g_0^2)$ of the off-diagonal components
of $T_{\mu\nu}$ in a broad range of values of the bare coupling $g_0^2$, between 0 and 1. This method is based on the 
framework of shifted boundary conditions~\cite{Giusti10,Giusti12} where one considers the definition of a thermal 
quantum field theory in a moving reference frame. Recently, the Wilson flow~\cite{Luscher09} has been suggested as 
an alternative method to compute the renormalization constants of the energy-momentum tensor~\cite{Suzuki13,DelDebbio13}.

\section{Renormalization of the  energy-momentum tensor}
In this section we present a method to evaluate the renormalization constants
of the traceless components of the energy-momentum tensor on the lattice. We discuss the case of the $SU(3)$
Yang-Mills theory but the method can be generalized in a straightforward way to a generic
gauge symmetry group. The gauge field $U_\mu (x) \in SU(3)$ is defined on the links of a
four dimensional lattice $L^3\times L_0$ and the interaction is described by the Wilson action
\begin{equation}
S[U] = -\frac{1}{g_0^2}  \sum_{x,\mu\nu} 
 \mbox{Re}\, \mbox{Tr} [U_\mu(x)U_\nu(x+\hat\mu)U^\dagger_\mu(x+\hat\nu) U^\dagger_\nu(x)]
\end{equation}
where $g_0$ is the bare coupling.
We impose periodic boundary conditions in the spatial directions and shifted boundary 
conditions along the temporal direction, 
$U_\mu(L_0,{\bf x}) = U_\mu(0,{\bf x}- L_0\, {\bf \xi})$,
where $(L_0/a)\, {\bf \xi}$ is a vector with integer components and $a$ is the lattice
spacing. The partition function is given by
\begin{equation}\label{partfun}
Z(L_0,{\bf \xi}) = \prod_{x,\mu} \int dU_\mu(x)\;  e^{-S[U]} = 
\mbox{Tr} [ e^{-L_0 (H + i {\bf \xi} \cdot {\bf P})}]
\end{equation}
where $H$ and $\bf P$ are the Hamiltonian and the total momentum operator, respectively.
We consider the clover definition of 
the energy-momentum tensor on the lattice~\cite{Caracciolo89}
\begin{equation}\label{eq:TmunuLat}
T_{\mu\nu} =  \frac{1}{g_0^2}\Big\{F^a_{\mu\alpha}F^a_{\nu\alpha}
- \frac{1}{4} \delta_{\mu\nu} F^a_{\alpha\beta}F^a_{\alpha\beta} \Big\}\; .
\end{equation}
The field strength tensor is defined as 
\begin{equation}
F^a_{\mu\nu}(x) = - \frac{i}{4 a^2} 
\mbox{Tr} \Big\{\Big[Q_{\mu\nu}(x) - Q_{\nu\mu}(x)\Big]T^a\Big\}\; ,  
\end{equation}
where $T^a=\lambda^a/2$ with $\lambda^a$ being the Gell-Mann matrices, and $Q_{\mu\nu}(x)$
is defined as follows
\begin{equation}
Q_{\mu\nu}(x) = P_{\mu\nu}(x) + P_{\nu-\mu}(x) + P_{-\mu-\nu}(x) +P_{-\nu\mu}(x)\; .
\end{equation}
The matrix $P_{\mu\nu}(x)$ is the parallel transport along an elementary plaquette at the
lattice site $x$ along the directions $\mu$ and $\nu$, and the minus sign stands for the
negative orientation. The diagonal and the off-diagonal components of the traceless 
part of the energy-momentum tensor renormalize
multiplicatively as~\cite{Caracciolo89} 
\begin{equation}\label{Zdef}
 T_{\mu\nu}^R  = Z_T(g_0^2)\;  T_{\mu\nu}\; , 
\qquad
T_{\mu\mu}^R - T_{\nu\nu}^R  = Z_d(g_0^2)\; ( T_{\mu\mu} - T_{\nu\nu} )\qquad \mbox{with} \qquad 
\mu\neq \nu\; , 
\end{equation}
where no summation is performed on the double indices $\mu\mu$ and $\nu\nu$. The
renormalization factors $Z_T(g_0^2)$  and $Z_d(g_0^2)$ depend on the bare coupling 
only. Their values at one loop in perturbation theory are~\cite{Caracciolo91}
\begin{equation}
Z_T(g_0^2) = 1 + 0.27076 \; g_0^2 + \ldots
\qquad \qquad
Z_d(g_0^2) = 1 + 0.24068\; g_0^2  + \ldots
\end{equation}
Since $\int d^3x T_{0k}$ is the charge of translation invariance, the expectation value of the 
renormalized operator $\langle T_{0k}^R \rangle$ can be can be directly obtained from eq.~(\ref{partfun}) 
\begin{equation}\label{T0kR}
\langle T_{0k}^R \rangle = \frac{1}{L^3L_0} \frac{\partial}{\partial \xi_k} \log
Z(L_0,{\bf \xi}).
\end{equation}
We can then use eq.~(\ref{Zdef}) and eq.~(\ref{T0kR}) to compute the renormalization factor
of the off-diagonal components of the energy momentum tensor as~\cite{Giusti14}
\begin{equation}\label{Z}
Z_T(g_0^2) = \frac{1}{L^3L_0} \frac{\frac{\partial}{\partial \xi_k} \log Z (L_0,{\bf \xi})}{\langle T_{0k} \rangle }
\end{equation}
Note that $\langle T_{0k} \rangle$ is measured using shifted boundary conditions with
shift $\bf \xi$.  A similar method, defined in the framework of shifted boundary
conditions, has been considered in~\cite{Robaina13}. In the limit of infinite spatial
volume, the renormalized space-time components of the energy-momentum tensor are also
related to the following combination of the diagonal components~\cite{Giusti12}
\begin{equation}
\langle T_{0k}^R \rangle = \frac{\xi_k}{1-\xi_k^2} \langle T_{00}^R - T_{kk}^R\rangle.
\end{equation}
Using eq.~(\ref{Zdef}), we can now evaluate the renormalization factor of the diagonal
components by
\begin{equation}\label{Zd}
z_T(g_0^2) = \frac{Z_d(g_0^2)}{Z_T(g_0^2)} = \frac{1-\xi_k^2}{\xi_k} \frac {\langle T_{0k}
  \rangle }{\langle T_{00} - T_{kk} \rangle }.
\end{equation}
This equation holds exactly also in finite volume for specific values of spatial lengths 
and shifts~\cite{Giusti12}.

\section{Numerical computation of $Z_T(g_0^2)$}
In this section we present the results of the numerical study to compute the
renormalization factor $Z_T(g_0^2)$ of the off-diagonal components of the energy-momentum
tensor using eq.~(\ref{Z}). We discretize the derivative and we write
\begin{equation}\label{Zpractic}
Z_T(g_0^2) = \frac{1}{2a L^3} \frac{\log Z(L_0,{\bf \xi }+a/L_0 \hat k ) -  
\log Z(L_0,{\bf \xi}-a/L_0 \hat k )}{\langle T_{0k} \rangle }\; . 
\end{equation}
As in any non-perturbative renormalization condition, the r.h.s. of the formula above has 
discretization effects. The corrections depending on 
$a/L$ and $a/L_0$ can be removed by taking the limits $L\rightarrow \infty$ and
$L_0\rightarrow \infty$. As we shall see, those corrections turn out to be very small.
The non-trivial part in applying eq.~(\ref{Zpractic}) is the measurement of the numerator: it
corresponds to measuring the ratio of two partition functions with different shifts at the
same value of $L_0$ and $g_0^2$. That calculation cannot be performed in 
a single Monte Carlo simulation due to the very poor overlap of the relevant phase space 
of the two integrals. In this case we have used the Monte Carlo procedure of 
Refs.~\cite{deForcrand00,DellaMorte08,Giusti10}. We consider a set of $(n+1)$ systems with action
$\overline S[U,r_i]= r_i S[U^{({\bf \xi} - a/L_0 \hat k)}] + (1-r_i) S[U^{({\bf \xi} + a/L_0 \hat k)}]$
($r_i=i/n$, $i=0,1,\dots,n$), where the superscript indicates the shift in the 
boundary conditions. The relevant phase space of two successive systems with $r_i$ and
$r_{i+1}$ is very similar and the ratio of their partition functions,
${\cal Z}(\beta,r_i)/{\cal Z}(\beta,r_{i+1})$, can be efficiently measured 
as the expectation value of the observable $O(U,r_{i+1}) = \exp{({\overline S}[U,r_{i+1}]-{\overline S}[U,r_i])}$
on the ensemble of gauge configurations generated with the action 
${\overline S}(U,r_{i+1})$. The discrete derivative  
is then written as 
\begin{equation}\label{eq:prod}
\frac{1}{2 a}\log {\frac{Z(L_0,{\bf \xi} + a/L_0 \hat k)}{Z(L_0,{\bf \xi} - a/L_0 \hat k)}}
 = \frac{1}{2 a} \sum_{i=0}^{n-1}
\log {\frac{{\cal Z}(\beta,r_i)}{{\cal Z}(\beta,r_{i+1})}}\; .
\end{equation}
The calculation of the r.h.s. becomes quickly demanding for large spatial volumes. We have
performed numerical simulations with $L=12$ and 16.

However, there is an alternative and more efficient method to compute the ratio of the two partition
functions. Calling $f(L,L_0,{\bf \xi},g_0^2)$ the l.h.s. of eq.~(\ref{eq:prod}), we rewrite it as 
\begin{equation}\label{intder}
f(L,L_0,{\bf \xi},g_0^2) = c_0 + \int_0^{g_0^2} \frac{\partial}{\partial x}
f(L,L_0,{\bf \xi},x) \; dx =c_0 +  \int_0^{g_0^2} \frac{dx}{x} 
(\langle S[U,{\bf \xi} + a/L_0 \hat k] \rangle - \langle S[U,{\bf \xi} - a/L_0 \hat k] \rangle)
\end{equation}
where $c_0 = f(L,L_0,{\bf \xi},0)$ is known analytically. Although the two
v.e.v.'s on the r.h.s. are fairly close, their difference can be computed at a few permille
accuracy with very moderate numerical resources. The integral is also
well-behaved around $g_0^2=0$: one can show that the difference of the two v.e.v.'s
vanishes at leading order in perturbation theory. Finally, it is important to notice that
the spatial size is no longer a problem in computing $f(L,L_0,{\bf \xi},g_0^2)$ with
eq.~(\ref{intder}). In fact, the increase of the computational effort due to larger
spatial volumes is compensated by the reduction of the statistical uncertainty in
the measurement of the two v.e.v.'s.

\begin{figure}[htb]
\begin{center}
\includegraphics[width=0.6\textwidth]{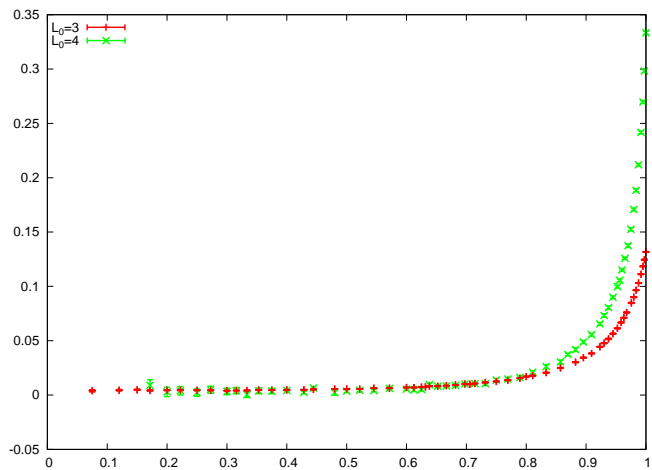}
\caption{The dependence of 
$(\langle S[U,{\bf \xi} + a/L_0 \hat k] \rangle - \langle S[U,{\bf \xi} - a/L_0 \hat k]
\rangle)/g_0^2$ on the bare coupling $g_0^2$.
The data have been produced for $L_0=3$ and $L_0=4$ with shift ${\bf \xi} = (1,0,0)$. }\label{numerator}
\end{center}
\end{figure}

\begin{figure}[htb]
\begin{center}
\includegraphics[width=0.6\textwidth]{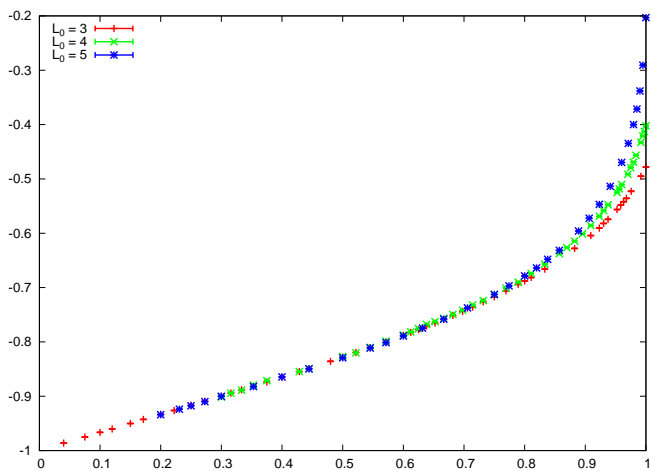}
\caption{The dependence of $\langle T_{0k}\rangle $ on the bare coupling $g_0^2$ for
  $L_0=3, 4$ and 5. The data are normalized by $|d_0|$, where $d_0$ is the value of
  $\langle T_{0k}\rangle $ at $g_0^2=0$. The value of the shift is ${\bf \xi} = (1,0,0)$.}\label{denominator}
\end{center}
\end{figure}

\begin{figure}[htb]
\begin{center}
\includegraphics[width=0.6\textwidth]{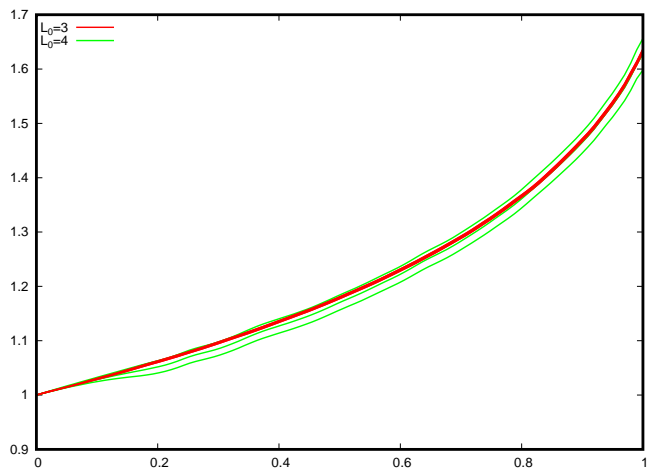}
\caption{The renormalization factor $Z_T(g_0^2)$ of the off-diagonal components of the
  energy-momentum tensor as a function of the bare coupling $g_0^2$.
The red and green data correspond to $L_0=3$ and 4, respectively.}\label{Zfinal}
\end{center}
\end{figure}

\begin{figure}[htb]
\begin{center}
\includegraphics[width=0.6\textwidth]{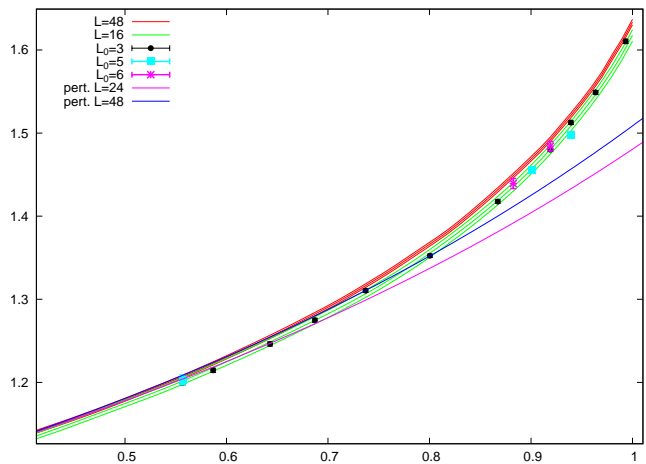}
\caption{Comparison of different methods to compute the renormalization factor
  $Z_T(g_0^2)$ as a function of the bare coupling $g_0^2$. The red and green curves are
  generated using eq.~(\protect\ref{intder}) on lattices with spatial size $L=48$ and 16,
  respectively, and temporal size $L_0=3$. The black, cyan and purple symbols are produced
  using eq.~(\protect\ref{eq:prod}) at $L=16$ and $L_0=3$, 5 and 6 respectively. The
  two-loop perturbative expansion are shown in pink and blue for $L=24$ and
  48.}\label{Zcompare}
\end{center}
\end{figure}

We have performed numerical simulations with shift ${\bf \xi} = (1,0,0)$ on lattices with
spatial size $L=48$. In figure~\ref{numerator}, we plot 
$(\langle S[U,{\bf \xi} + a/L_0 \hat k] \rangle - \langle S[U,{\bf \xi} - a/L_0 \hat k] \rangle)/g_0^2$
as a function of $g_0^2$. Red and green symbols refer to $L_0=3$ and $L_0=4$,
respectively. Monte Carlo simulations are in progress for $L_0=5$. 
In figure~\ref{denominator} we show the dependence of $\langle T_{0k} \rangle$ on $g_0^2$; the
data have been normalized by $|d_0|$, where $d_0$ is the value of $\langle T_{0k} \rangle$
at $g_0^2=0$ and it has been computed analytically~\cite{Giusti12}.
The integral of eq.~(\ref{intder}) is performed by numerical integration and then, by taking
the ratio with $\langle T_{0k}\rangle $, one can obtain the dependence of $Z_T(g_0^2)$ on $g_0^2$. 
The data are shifted by $-c_0/d_0+1$ in order to reduce the corrections in
$a/L$ and $a/L_0$. The results are plotted in figure~\ref{Zfinal} and show almost no
dependence on $L_0$; also the discretization effects in $L$ are smaller than the statistical
errors.  
In order to have a check of the reliability of the method, we have performed the above
described calculation of $Z_T(g_0^2)$ on a lattice with spatial size $L=16$ and
$L_0=3$. These data -- shown using the red and green lines in figure~\ref{Zcompare} -- can
be directly compared with those produced at $L=16$ using eq.~(\ref{eq:prod}) which are
plotted with the black symbols. The cyan and purple symbols correspond to data still
obtained with eq.~(\ref{eq:prod}) at $L=16$ but with $L_0=5$ and 6, respectively. These
two sets show that the dependence on $a/L_0$ is not visible within the numerical
accuracy. Finally, as a further check, we have computed the perturbative expansion of $Z_T(g_0^2)$ at
two loops using the method of Numerical Stochastic Perturbation Theory~\cite{DiRenzo94}
at $L=24$ and 48. These latter results, show evidence both for strong finite size effects
and for large corrections due to high-order terms and to non-perturbative contributions.

\section{Conclusions}
In this talk we have presented preliminary results for the computation of the renormalization factor of the
energy-momentum tensor in $SU(3)$ Yang-Mills theory. We propose a method that allows to attain
an accuracy of a few permille in a broad range of values of the bare coupling $g_0^2$, between 0 and 1 with a 
moderate numerical effort. The calculation of the renormalization factor is an important input for extracting 
physically relevant information from the energy-momentum tensor in a Monte Carlo simulation on the lattice.

\end{document}